\documentclass[12pt, preprint]{aastex}
\usepackage{float}
\usepackage{amsmath}
\usepackage{epsfig,floatflt}
\usepackage{subfigure}

\begin{document}

\title{Bayesian Power Spectrum Analysis of the First-Year \emph{WMAP} data }

\author{I.\ J.\ O'Dwyer\altaffilmark{1},
H.\ K.\ Eriksen\altaffilmark{2,3,4,5}, 
B.\ D.\ Wandelt\altaffilmark{1,6},
J.\ B.\ Jewell\altaffilmark{4},
D.\ L.\ Larson\altaffilmark{6},
K.\ M.\ G\'orski\altaffilmark{4,5,7},
A.\ J.\ Banday\altaffilmark{8},
S.\ Levin\altaffilmark{4},
P.\ B.\ Lilje\altaffilmark{2}
} 

\altaffiltext{1}{Astronomy Department,
    University of Illinois at Urbana-Champaign, IL 61801-3080}
\altaffiltext{2}{Institute of Theoretical Astrophysics, University of Oslo,
  P.O.\ Box 1029 Blindern, N-0315 Oslo, Norway}
\altaffiltext{3}{Centre of Mathematics for Applications,
University of Oslo, P.O.\ Box 1053 Blindern, N-0316 Oslo}
\altaffiltext{4}{JPL, M/S 169/327, 4800 Oak Grove Drive,
  Pasadena CA 91109} 
\altaffiltext{5}{California Institute of Technology, Pasadena, CA
  91125}
\altaffiltext{6}{Department of Physics, University of Illinois at Urbana-Champaign, IL 61801-3080}
\altaffiltext{7}{Warsaw University Observatory, Aleje Ujazdowskie 4,
  00-478 Warszawa, Poland}
\altaffiltext{8}{Max-Planck-Institut f\"ur Astrophysik, Karl-Schwarzschild-Str.\
1, Postfach 1317, D-85741 Garching bei M\"unchen, Germany}


\begin{abstract}

We present the first results from a Bayesian analysis of the \emph{WMAP} 
first year data using a Gibbs sampling technique. Using two independent, 
parallel
supercomputer codes we analyze the \emph{WMAP} Q, V and W bands. The
analysis results in a full probabilistic description of the information
the \emph{WMAP} data set contains about the power spectrum and the
all-sky map of the cosmic microwave background anisotropies. We present the 
complete probability distributions for
each $C_\ell$ including any non-Gaussianities of the power spectrum
likelihood. 
While we find good overall agreement with the previously published 
\emph{WMAP} spectrum, our analysis uncovers discrepancies in the power 
spectrum estimates at low $\ell$ multipoles. For example we claim the 
best-fit $\Lambda$CDM model is consistent with the $C_2$ inferred from our 
combined Q+V+W analysis with a 10\% probability of an even larger 
theoretical $C_2$. Based on our exact analysis we can therefore attribute 
the "low quadrupole issue" to a statistical fluctuation.
\end{abstract}



\section{Introduction}

Cosmic Microwave Background (CMB) power spectrum estimation from large
datasets such as the Wilkinson Microwave Anisotropy Probe (\emph{WMAP})
\citep{bennett:2003a} presents a considerable computational 
challenge.   With the exception of some specialized methods \citep{oh:1999,WH2003}, obtaining 
the power spectrum and error bars for a generalized, large CMB 
dataset without introducing significant simplifications has, 
until now, been computationally impossible.  However, a 
new numerical approach based on Gibbs sampling has been developed recently by   
Jewell, Levin, and Anderson (2004) and Wandelt, Larson, and
Lakshminarayanan (2004) which offers the hope of  
overcoming these computational difficulties and allowing the analysis of
large CMB datasets regardless of scanning strategy, noise characteristics 
and so on.  In this letter we present our results from the application of 
this new method to the first year \emph{WMAP} data.  We developed two
independent, parallel codes to achieve this and a detailed description of the 
implementation of these codes is given in a
companion publication \citep{eriksen:2004}. The analysis
results in a full, multivariate probabilistic description of the information the
\emph{WMAP} data set contains about the power spectrum and the all-sky map of the
CMB anisotropies.  For the purposes  of this letter we limit ourselves to a presentation of the
results at each angular scale $\ell$,   leaving a full multivariate
exploration of the \emph{WMAP} likelihood  for a
future publication. In \S 2 we give a very brief 
overview of our method and in \S 3 we present on overview of the
implementation of our codes.  \S 4 contains our results, provides 
some interpretation and discusses what light our method can
shed on some of the anomalies in the \emph{WMAP} power spectrum that have been
discussed in the literature (e.g.~\citet{hinshaw:2003,
  spergel:2003,efstathiou:2004, slosar:2004}).  Our conclusions and future
  directions for this work are detailed in \S 5. 

\section{Method Overview}

In this section we present only a very brief overview of our method and 
codes and refer the reader to \citet{eriksen:2004} for a detailed discussion
of our implementations and \citet{jewell:2002} and \citet{wandelt:2003} for
a discussion of the underlying principles.

The main difference between our approach and previous attempts at power 
spectrum estimation is that rather than trying to solve the maximum likelihood 
problem directly, we \emph{sample} from the probability density of the power
spectrum $C_\ell$ given the data $\mathbf{d}$,
$P(C_{\ell} | \mathbf{d})$.  While directly sampling from this density is
difficult or impossible, we can sample from the joint posterior
density $P(C_{\ell}, \mathbf{s} | \mathbf{d})$, where $\mathbf{s}$ 
is the CMB signal map. Expectation values of any statistic of the
$C_{\ell}$ will converge to the expectation values of
$P(C_{\ell} | \mathbf{d})$.  The theory of Gibbs sampling \citep{gelfandsmith:1990, tanner:1996} tells us that if we can 
sample from the conditional densities $P(\mathbf{s}|C_{\ell}, \mathbf{d})$ and
$P(C_{\ell}|\mathbf{s},\mathbf{d})$ then iterating the  following two sampling equations
will, after an initial burn-in period, lead to a sample from 
the joint posterior  $P(C_{\ell}, \mathbf{s}| \mathbf{d})$:
\begin{align}
\mathbf{s}^{i+1} &\hookleftarrow P(\mathbf{s}|C_{\ell}^{i}, \mathbf{d}), \\
C_{\ell}^{i+1} &\hookleftarrow P(C_{\ell}|\mathbf{s}^{i+1}).
\end{align}
Here the symbol ``$\hookleftarrow$'' denotes drawing a random realization from the
density on the right.  The conditional density of the signal given the
most recent $C_{\ell}$ sample is $P(\mathbf{s}|C_{\ell}^{i},
\mathbf{d})\propto G(S^i(S^i+N)^{-1}m,((S^i)^{-1}+N^{-1})^{-1})$,
where $\mathbf{m}$ is the map constructed from the data $\mathbf{d}$.  To
sample from this density we simply generate a Gaussian variate with 
the required mean and covariance.  The density for the $C_{\ell}$
factorizes to an inverse gamma distribution due to the special form of
$S$ and sampling from this is very simple.  For each $\ell$ we compute
$\sigma_{\ell}=\sum^{+l}_{m={-l}}|s^i_{lm}|^2$ and a (2$\ell$-1)-vector
of Gaussian random variates with zero mean and unit variance.  Then
$C_{\ell}^{i+1}={\sigma_{\ell}\over{|\rho_l|^2}}$.  Given this sample from 
$P(C_{\ell}, \mathbf{s}|\mathbf{d})$, the marginalization over 
$\mathbf{s}$ to obtain $P(C_{\ell}|\mathbf{d})$ is trivially
accomplished: just ignore the
$\mathbf{s}$ in the tuple $(C_\ell,\mathbf{s})$. Of course this is not to say
that the $\mathbf{s}$ sample is purely ancillary in the analysis: it is also of interest to
explore $P(\mathbf{s}|\mathbf{d})$ which is a full representation of the
CMB signal content of the data.

The scaling of this Monte Carlo method is
$\mathcal{O}(N_{\textrm{pix}}^{3/2})$ and corresponds to the scaling of
the spherical harmonic transforms performed by the HEALPix algorithm.  
Other issues which affect our
computational cost are the choice of a good preconditioner for our linear
algebra solver, the correlation length between samples and the length of
burn-in time required for the sampling scheme, eqs.~1 and 2, to converge.  We
perform the sampling in eq.~1, by means of solving linear systems of equations
using the Conjugate Gradient algorithm. The computational cost of this
method is therefore highly dependent on the choice of a good preconditioner
for this system.  We experimented with several preconditioners and found
examples which gave convergence in a few tens of iterations.  The number of
samples we need to obtain in order to adequately describe the statistics of
the power spectrum will depend upon the degree of correlation between
successive samples.  A detailed discussion of these issues can be found in
\citet{eriksen:2004}.  


\subsection{Foregrounds}
An appealing feature of the sampling approach is its ability to incorporate
virtually any real-world complications, as discussed by \citet{jewell:2002}
and \citet{wandelt:2003}.  A few examples of this flexibility are applications
to $f^{-1}$ noise, asymmetric beams or arbitrary sky coverage.  Foreground
estimation and removal are also possible.  In principle, detailed prior
knowledge of all anticipated foregrounds can be included and the algorithm
will then return the level of the foreground supported by the data in the
map. However, for this first analysis we limit ourselves to including two
simple foreground components: (1) a stochastic model of the monopole and
dipole in the maps, and (2) a stochastic model of the foreground
component. These models are encoded in terms of uniform, improper priors that express
our ignorance of the monopole and dipole in the maps as well as the foreground
contributions in the regions that are flagged in the foreground masks supplied
by the \emph{WMAP} team. Traditionally this masked region is excluded from the
analysis altogether. In the Bayesian approach it is modeled as a region where
the foregrounds are completely unknown. Sampling from the posterior density
reconstructs the signal in the masked region, based on the correlation
structure discovered on the unmasked portion of the sky.  The details of these
foreground models are outlined in the above papers as well as in
\citet{eriksen:2004}.

\section{Implementation}

We used two independently developed implementations of the algorithm to 
produce our results:  MAGIC \citep{wandeltconf:2004}  and Commander.  
These are both  parallel implementations of the sampling algorithm. Having 
results from both MAGIC and Commander allowed us to cross-check results and 
provided valuable redundancy. Our tests of the codes are detailed in
\citet{eriksen:2004} and these showed that both codes were producing
results consistent with each other. In both codes our noise model for
each channel consisted of combining the published  
number of observations per pixel with the noise per sample for each channel
and assuming the noise to be uncorrelated.  We also show in 
\citet{eriksen:2004}  that residual correlated noise in the \emph{WMAP} maps 
has negligible effect on our results. 

The data input to these codes were the first year \emph{WMAP}
maps\footnote{available at http://lambda.gsfc.nasa.gov}  for
all 8 of the cosmologically interesting frequency bands (2 at Q-, 2 at V- 
and 4 at W-band). For all channels we used  template corrected maps following \citet{bennett:2003a}.
We performed two analyses with these  data. First we analyzed the data
in the  V and W channels separately by band, using
  unweighted averages of the channel maps and computing an effective beam window
function. For these runs we chose the conservative Kp0 mask and ran single,
long chains, containing 1000 samples for each band.  Q-band was
analyzed more aggressively, combining the channels on
the likelihood level \citep{eriksen:2004}, and using the Kp2 mask to
assess potential foreground contamination.   These 
analyses will be labeled Q, V and W in the following.

The second  analysis uses the optimal combination of all 8
channels in the likelihood using their respective 
noise specifications and beam window functions.   We will refer to this joint
analysis as QVW in the following.
As before, we specified
a stochastic model for the monopole and dipole component as well as
complete ignorance about the foregrounds within a  region on the
sky. For the QVW analysis this region was defined by the more aggressive Kp2 mask.
Both the Q and QVW  analyses were done in parallel, short
chains ($\sim$10 chains, $\sim$100 samples each) initialized with plausible
starting spectra inspired by the \emph{WMAP} analysis
\citep{hinshaw:2003}. 
 
\section{Results}  All of the results presented here are based on $\sim$ 1000 samples for each
frequency band, which required a total of $\sim$15,000 hours of CPU time 

Figure \ref{fig:QVWsignal} shows the Bayes estimate (posterior mean) for the
  signal QVW analysis. This map represents the CMB signal content in the Q, V,
  and W bands of the \emph{WMAP}
  data. This map shows detail where the data warrants
  it and smoothes where the data is poor. The CMB signal is clearly visible 
outside the Galaxy and the algorithm has reconstructed the signal in
the Galactic plane for the largest scale modes, consistent with the
signal phases and correlations it has discovered on the high latitude
sky. There is insufficient information in the map for the algorithm
to reconstruct the smaller scale modes in the galactic plane. 
This result is a non-linear generalization of the 
Wiener filter which does not assume prior information
about the signal correlations but discovers them from the data. 

Figure \ref{fig:4ps} shows the power spectra we obtain for Q-,
V- and W-band and the combined QVW analysis.  Large scale features in
the power spectra are consistently reproduced across all the channels.

In Figure \ref{fig:20panelcomb}, we show the probability distribution of the
$C_\ell$ samples for some $\ell$ of interest.  Since the original 
\emph{WMAP} analysis was released there has been some controversy
  regarding a lack of power at the largest angular scales (lowest $\ell$'s)
  e.g. \citep{efstathiou:2004, slosar:2004}.   It can be
seen from the plots that our estimates of these $C_{\ell}$'s are consistent
with the original determinations made by the \emph{WMAP} team, but
that there are significant, non-Gaussian uncertainties on our estimates which indicate 
that low values of $C_\ell$ for the largest angular scales are not so
unexpected.  We also over-plot the
results using the \emph{WMAP} internal linear combination map with the
Kp2 mask from \citet{efstathiou:2004} in our figure.  It can be seen
that there is good agreement between the values obtained in our
analysis and that of Efstathiou.

For easy reference we show the cumulative probability distributions for the
lowest $\ell$ in Figure
\ref{fig:odds}. One can read off the probability that the actual theory
$C_\ell$ are lower than the prediction from any particular model. Probabilities
close to 0 or 1 indicate a poor fit, or outlier. A search over all strongly signal dominated
$\ell$ from 2 to 350 using our V and W band analyses resulted
in  10 outliers (within 0.01 of 0 or 1) for V band ($\ell=$54, 73, 114, 117, 121, 179, 181, 209, 300,
322) and 9 outliers for W band ($\ell$=73, 82, 117, 121, 181, 261, 334, 341,
344). For 349 tests this corresponds to only a slightly higher number of outliers
than expected based on counting statistics, and is entirely unremarkable if we
only count those outliers that appear in both bands. Based on
this $\ell$-by-$\ell$ analysis the ``bite'' in the spectrum does not seem
particularly extraordinary, with only one outlier in the range of
200-220. However, all $C_\ell$ at $205<\ell<210$ are quite far below the best fit and will most likely
be noted as a highly significant outlier in a full multivariate analysis of the joint
posterior density. 

\section{Discussions and Conclusions}
\label{sec:conclusions}

We have implemented the Gibbs sampling technique introduced by
\citet{jewell:2002} and \citet{wandelt:2003}, and applied it to 
the \emph{WMAP} data.  This first presentation of our results focuses on the
inferred marginalized likelihoods for each $\ell$. We find that these are
broadly consistent with previous  
determinations of the power spectrum. By
virtue of using a Bayesian analysis we can present 
present a more complete picture of the
uncertainties in the estimates across all relevant scales. Previous likelihood-based work to generate a full
probabilistic description (e.g.~\citet{slosar:2004}) were limited to looking at
only the lowest $\ell$ due to the computational difficulty involved. We present
an analysis that 
covers the region from $\ell=2$ to $\ell=500$. A multivariate treatment will 
therefore allows a rigorous assessment of  
the significance of anomalies in the power spectrum such as the low power
on large angular scales, the ``bite'' in the power spectrum near the peak,
etc., as well as using the full multivariate posterior density as the input to
parameter estimation. This is particularly interesting since the Bayesian
approach yields a \emph{converging} analytic approximation to 
the likelihood of $C_\ell$ given the data, it is not necessary to
construct parametric approximations to the likelihood such as the Gaussian or shifted log-normal
approximations. A detailed study of these issues will be presented in a future
publication. 

More  information can be drawn from the data
in the future.  The extension 
to more complex foreground models, the inclusion of correlated noise and 
the addition of polarization to the analysis will all be of great interest, 
especially in view of the upcoming release of the second year of 
\emph{WMAP} data and, further in the future, the \emph{Planck} experiment.

\begin{acknowledgements}
We acknowledge use of the
HEALPix\footnote{http://www.eso.org/science/healpix/} software (G\'orski, Hivon \& Wandelt 1998) and analysis package.  We also acknowledge use of the Legacy
Archive for Microwave Background Data Analysis (LAMBDA).  H.\ K.\ E.\
and P.\ B.\ L.\ acknowledge financial support from the Research
Council of Norway, including a Ph.\ D.\ studentship for H.\ K.\
E.  This work has also received support from The Research Council of
Norway (Programme for Supercomputing) through a grant of computing
time.  This work was partially performed at the Jet Propulsion
Laboratory, California Institute of Technology, under a contract with
the National Aeronautics and Space Administration. This work was partially
supported by an NCSA Faculty Fellowship for B.D.W.  This research 
used resources of the National Energy Research Scientific Computing
Center, which is supported by the Office of Science of the
U.S. Department of Energy under Contract No. DE-AC03-76SF00098. 

\end{acknowledgements}

\clearpage

\figcaption{The posterior mean $\langle\mathbf{s}\rangle_{P(\mathbf{s}| \mathbf{d})}$ of the
  CMB signal $\mathbf{s}$ given the \emph{WMAP} first year data, including all
  8 channels in the Q, V and W   bands. This map represents the CMB
  signal content of the \emph{WMAP} data. It
  contains detail where the data warrants it and smoothes where the data is
  poor.   For example, the algorithm is able to reconstruct the
  largest scale modes in the Galactic plane based on the signal phases
  and correlations it discovers in regions of the map outside the
  Galactic plane. The smallest scale which can be reproduced in the Galactic
  plane is set by the mask size and is a natural consequence of Wiener
  filtering. 
\label{fig:QVWsignal}}

\figcaption{Four power spectra obtained from the Gibbs sampling
  algorithm.  The uppermost panel shows the Q-band power spectrum,
  then V-band, W-band and finally the lower panel the combined
  analysis from all eight cosmologically interesting \emph{WMAP}
  channels.  Gray dots represent the individual samples and the
  dark line is the mode of the unbinned power spectrum.  The lighter gray
  line in the bottom panel is the \emph{WMAP} combined CMB power spectrum 
  and where the line is not visible the difference between our result and
  the original \emph{WMAP} result differ by less than the width of the
  line.  The Q and Q+V+W result was obtained with an initial power
  spectrum based on the \emph{WMAP} best-fit power spectrum with a random
  component, while the V and W results were obtained using an initial
  spectrum proportional to 1/${\ell(\ell+1)}$.  Our V and W results
  were obtained using the conservative Kp0 mask while our Q-band
  result uses the Kp2 mask.  This was done to assess potential foreground
  contamination, which we find no evidence for. 
\label{fig:4ps}}

\figcaption{The probability distributions of the $C_{\ell}$ samples for a selected set of
  $\ell$ giving the probability of obtaining a value 
of $C_{\ell}$ within a given histogram bin.  We plot results for the
  significantly non-Gaussian low-$\ell$ multipoles and for selected
  higher values based on their deviation from the best-fit
  $\Lambda$-CDM model.  There are 50 bins in each
  histogram. Red, green and blue
histograms are Q, V and W-band
 respectively.  Black is the combined QVW analysis. The dotted
 vertical line is the \emph{WMAP} best estimate of the $C_\ell$ value.
The solid line is the \emph{WMAP} best fit theory $C_\ell$'s and the 
dash-dot-dot line is the average of the samples from our algorithms.
For $\ell < 7$ the dashed lines show the \emph{WMAP}-ILC values from
 \citet{efstathiou:2004} for comparison.  Further plots are
  available at http://www.astro.uiuc.edu/$\sim$iodwyer/research\#wmap.
\label{fig:20panelcomb}}

\figcaption{Easy-to-use representation for evaluating theoretical models of the
  power spectrum at the
  lowest $\ell$. We plot the  probability $P(C_\ell^\mathrm{theory}<C_\ell)$
  against   $ \ell(\ell+1)C_\ell/2\pi$ for the largest angular scales probed by
 \emph{WMAP}. Theorists can evaluate the goodness of fit of their model $C_\ell$ with the
 \emph{WMAP} data by reading off the probability from this graph. If $P$ is within
  0.025 of 1 or 0, the model would be ruled out at the 5\% level based on that
  $\ell$ alone. The vertical dotted line is the \emph{WMAP} best fit theory assuming a constant scalar
  spectral index $n_s$. The probability is $\sim$90\% that the the actual theory
  $C_2$ is smaller.
\label{fig:odds}}

\clearpage

\pagestyle{empty}

\begin{figure}
\plotone{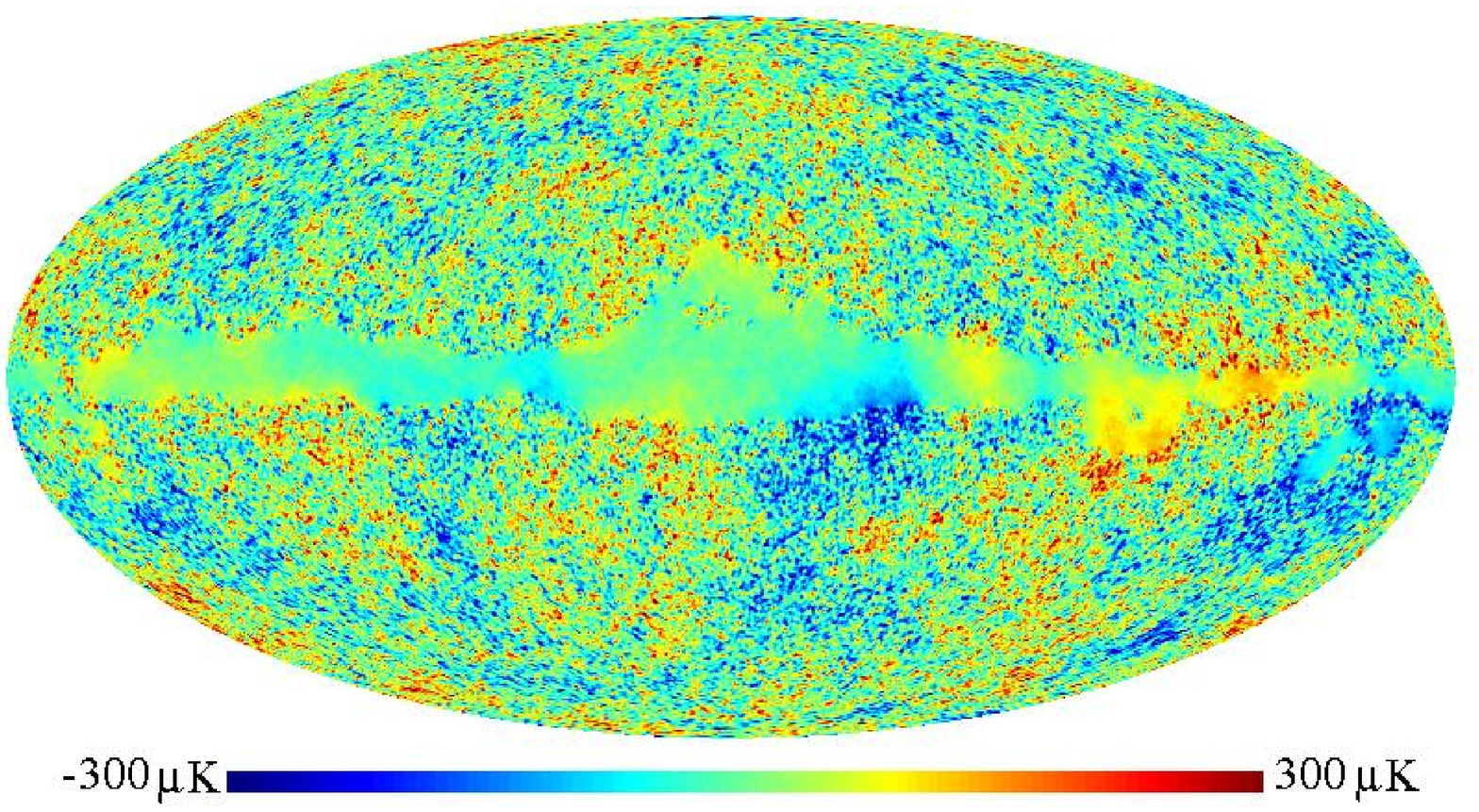}
\end{figure}

\begin{figure}
\vspace{-2cm}
\epsscale{0.9}
\plotone{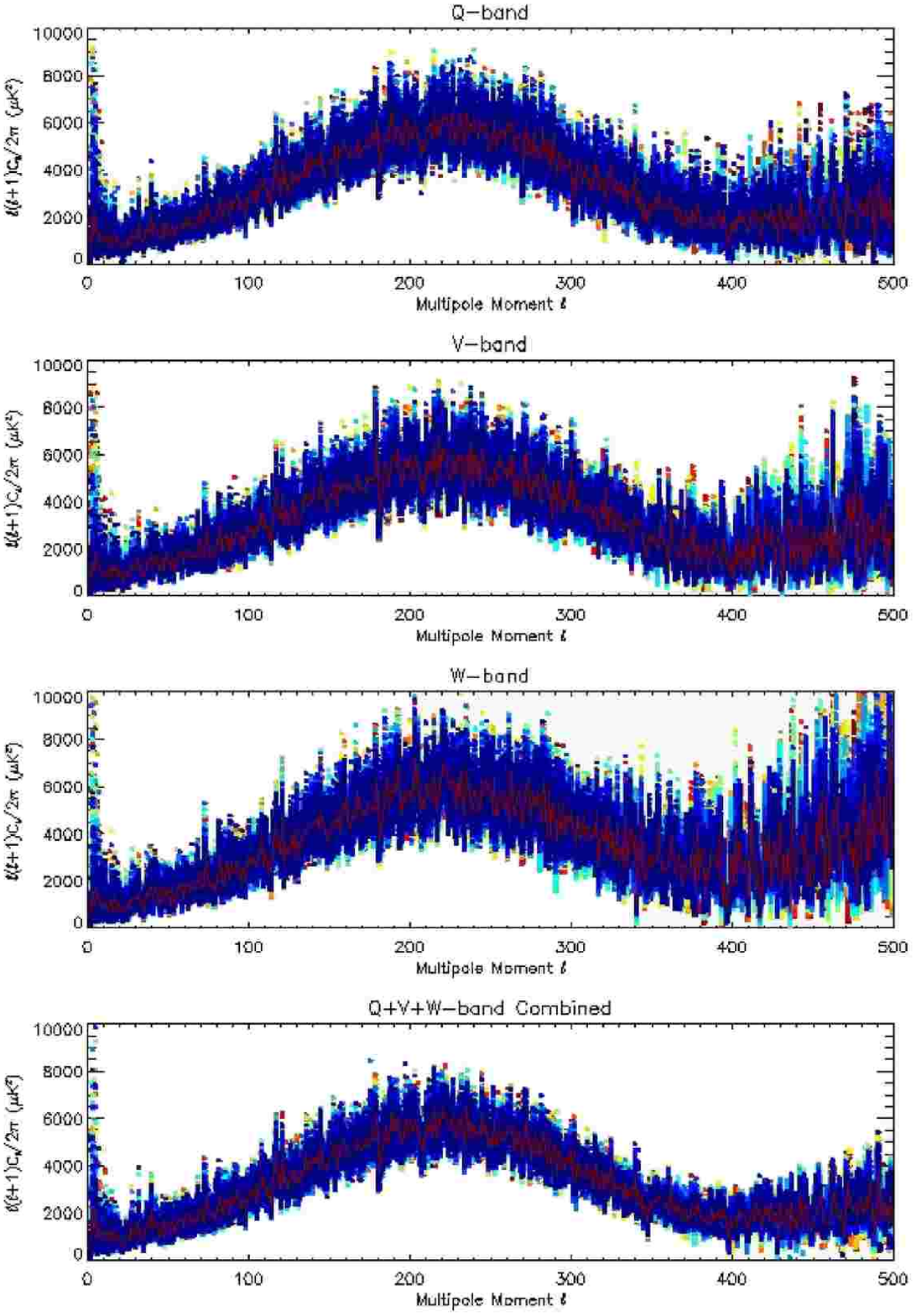}
\end{figure}

\begin{figure}
\epsscale{1}
\plotone{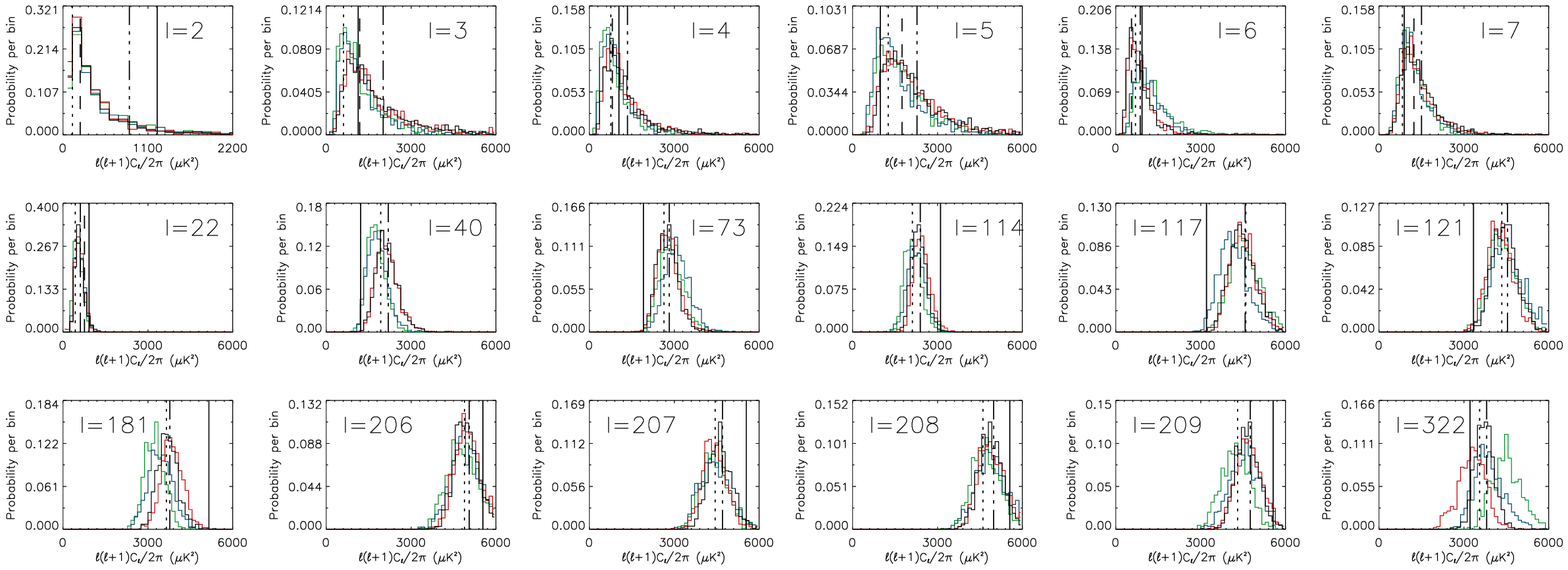}
\end{figure}

\clearpage

\begin{figure}
\plotone{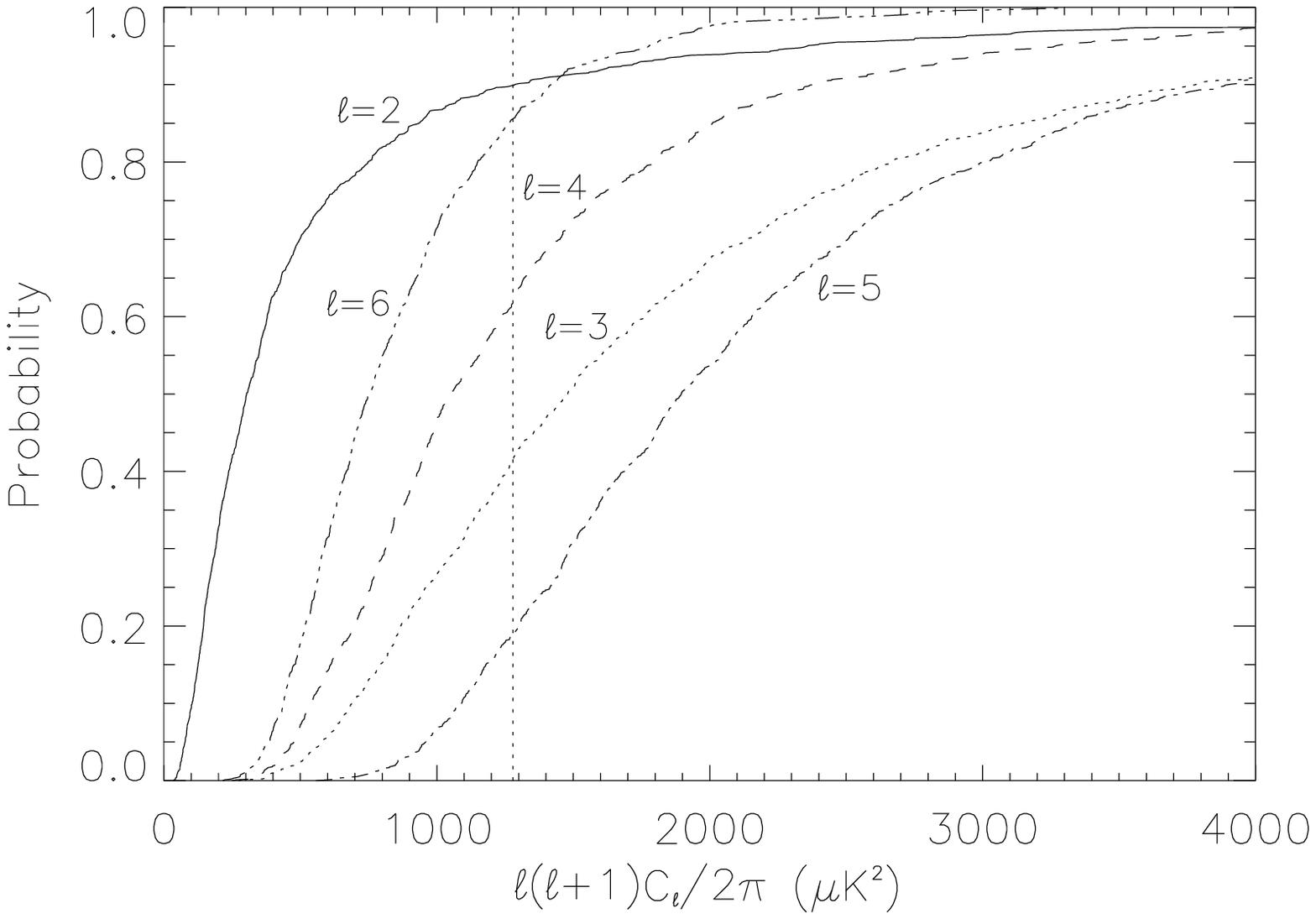}
\end{figure}

\end{document}